\documentclass[12pt]{article}
 
\usepackage{amssymb}

 \newcommand{\n}{\noindent} 
\newcommand{\rf}[1]{(\ref{#1})}
\newcommand{\ba}{\begin{array}} \newcommand{\ea}{\end{array}}
\newcommand{\be}{\begin{equation}} 
\newcommand{\btb}{\begin{tabular}}\newcommand{\etb}{\end{tabular}}
\newcommand{\ee}[1]{\label{#1}\end{equation}}
\newcommand{\bi}{\bibitem} 
\newtheorem{thm}{Theorem}[section] \newtheorem{pro}[thm]{Proposition}
\newtheorem{cor}[thm]{Corollary}\newtheorem{lem}[thm]{Lemma}
\newtheorem{df}[thm]{Definition} 

\newcommand{\bfl}{\begin{flushleft}}\newcommand{\efl}{\end{flushleft}}

 \newcommand{\G}{\Gamma}  
\newcommand{\De}{\Delta}\newcommand{\ep}{\epsilon}
\newcommand{\te}{\theta}\newcommand{\Te}{\Theta} \newcommand{\la}{\lambda}\newcommand{\La}{\Lambda}  
\newcommand{\Si}{\Sigma}

\newcommand{\LI}{\mathbb L}  \newcommand{\R}{\mathbb R} 
\newcommand{\N}{\mathbb N} \newcommand{\Q}{\mathbb Q}\newcommand{\Z}{\mathbb Z}  

\newcommand{\EC}{{\cal E}}
\newcommand{\GC}{{\cal G}}
\newcommand{\PC}{{\cal P}} 
\newcommand{\TC}{{\cal T}}
\newcommand{\LC}{{\cal L}}\newcommand{\LCU}{{\cal L}_\uparrow} 
\newcommand{\PCU}{{\cal P}_\uparrow}

\newcommand{\ay}{{\bf a}}\newcommand{\by}{{\bf b}}

\newcommand{\ky}{{\bf k}} 
 
\newcommand{\ry}{{\bf r}} 
\newcommand{\vy}{{\bf v}} 
\newcommand{\0}{{\mathbf 0}} \newcommand{\x}{{\bf x}} 
\newcommand{\Wy}{{\bf W}}

\newcommand{\sd}{\rtimes}

\newcommand{\hG}{\hat{G}}

 \newcommand{\hc}{\hat{c}}

 \newcommand{\we}{\wedge} 
\newcommand{\ra}{\rightarrow} \newcommand{\ol}{\overline} 
 
\newcommand{\sq}{\subseteq}\newcommand{\sqq}{\supseteq} 
\newcommand{\tl}{\tilde} 
\newcommand{\tlH}{\tilde{H}}

\newcommand{\Llr}{\Longleftrightarrow}

\begin{document} 

\title{Simultaneity as an Invariant Equivalence Relation}
\author{Marco Mamone-Capria\\ \small Dipartimento di
Matematica -- via Vanvitelli, 1 -- 06123 Perugia - Italy \\ \small {\sl E-mail}:
\texttt{mamone@dmi.unipg.it} }
\maketitle

\small
{\bf Abstract} This paper deals with the concept of simultaneity in classical and relativistic physics as construed in terms of  group-invariant equivalence relations. A full examination of Newton, Galilei and Poincar\'e invariant equivalence relations in $\R^4$ is presented, which provides alternative proofs, additions and occasionally corrections of results in the literature, including Malament's theorem and some of its variants. It is argued that the interpretation of simultaneity as an invariant equivalence relation, although interesting for its own sake, does not cut in the debate concerning the conventionality of simultaneity in special relativity. 

\normalsize

\tableofcontents

\section{Introduction}

In his electrodynamical paper of 1905 Einstein stated that ``all our statements in which time plays a role are always statements about simultaneous events'' (\cite{ei05}, p. 893), which implies that the time ordering of events presupposes the existence of a simultaneity relation between events. Such a binary relation quite naturally must be reflexive, symmetric and transitive, i. e. an equivalence relation on the space of events.\footnote{This is needed if we wish to build a time coordinate out of simultaneity, as in the approach to the Lorentz transformation adopted by Poincar\'e and Einstein.}  Now the usual statement that `there is no absolute simultaneity in special relativity', which has been made by Einstein himself, in an even blunter form,\footnote{In 1949, as the first of two ``insights of definite nature which physics owes to the special theory of relativity'', Einstein stated : ``There is no such thing as simultaneity of distant events; [...]'' (\cite{ei49}, p. 61).} has been construed by some authors as meaning that there is no nontrivial Poincar\'e-invariant equivalence relation on $\R^4$. Thus if we want to get a nontrivial equivalence relation, we must pass to a subgroup of the Poincar\'e group. By using the subgroup of all the Poincar\'e transformations fixing a pencil of parallel timelike worldlines (that is, the isotropy subgroup for a fixed notion of rest in Minkowski space-time), the only invariant relation we are left with is indeed standard simultaneity. A different subgroup is at the core of a well-known result by D. Malament published in 1977 (\cite{m77}).

In \cite{mm} I have argued at length, against conventionalists like Hans Reichenbach (\cite{re69}, \cite{re57}; cf. \cite{ja10}), that standard frame-dependent simultaneity does in fact deserve a privileged place within special relativity, and that the notion of `convention' used to deny this claim is methodologically inadequate. In \cite{mm} I did not discuss Malament's theorem, because I did not share the enthusiasm expressed by some authors. For instance, J. Norton in 1992 (cf. \cite{ss99}, p. 210) wrote that Malament's theorem had brought about ``one of the most dramatic reversals in debates in the philosophy of science'' (away from Reichenbachian conventionalism, that is). In this paper I deal in detail with this issue and explain the reasons behind my negative assessment.   

My treatment provides a unified presentation, alternative proofs, additional results and occasionally corrections, with respect to the papers by Malament (\cite{m77}),  Sarkar and Stachel (\cite{ss99}), Giulini (\cite{g}), and others. I conclude that the interpretation of simultaneity as an invariant equivalence relation, although interesting for its own sake, does not cut in the debate concerning the conventionality of simultaneity in special relativity. A different approach is necessary to refute this version of conventionalism.\footnote{As regards the conventionality issue, I stand by the results and conclusion of \cite{mm}, although I hope to come back to it, for commentary and extensions, in a future contribution.}    

\section{Invariant equivalence relations}

As explained in the introduction, one may explicate one aspect of the problem of simultaneity in special relativity as a request to classify invariant equivalence relations in Minkowski space-time with respect to suitable subgroups of the Poincar\'e group. Let us briefly remember first a few standard definitions and results.

\subsection{Basic definitions and notation}

An action of a group $G$ to the left on $X$ will be denoted by a dot: $g\cdot p$ is the effect of group element $g$ on point $p$. The orbit of a point $p$ is denoted by $G\cdot p$. An action is called {\sl transitive} if the orbit of some point (and therefore of every point) is the whole set $X$. 

For any $p\in X$ the set \( H(p):= \{g\in G \; : \; g\cdot p = p\}\) is easily seen to be a subgroup, and is called the {\sl isotropy subgroup} of $p$. For every $g\in G$ one has:

\be H(g\cdot p) = g H(p) g^{-1}.\ee{iso}

\n
so {\sl if the action is transitive}, all isotropy subgroups are isomorphic (indeed, conjugate) to each other.

An action is {\sl trivial} if, for every $p\in X$, $H(p)=G$ (i.e. $g\cdot p = p$ for all $g\in G$ and $p\in X$). 

An equivalence relation $\sim$ on $X$ is {\sl trivial} if it coincides with either $T:= X\times X$ (the ``total'' equivalence relation) or $I:=\De (X)$, the diagonal of $X$ (the ``identical'' equivalence relation).  For any given equivalence relation I shall denote by the symbol $[p]$ the equivalence class of any point $p$.

\subsection{Lattices of equivalence relations on a set}

The set of all equivalence relations $\EC (X)$ on $X$ is partially ordered by declaring a relation $R$ to be {\sl finer than} $R'$ (or \( R \preceq R'\)) when $R \subseteq  R'$, or equivalently when for all $p\in X$  \( [p]_{R'}\supseteq [p]_R \). Clearly the total and the identical relation, $T$ and $I$, are respectively the absolute maximum and the absolute minimum of this set. Moreover, for any two elements $R_1, R_2$ in $\EC (X)$ there exists the lowest upper bound and the highest lower bound, denoted by $R_1 \vee R_2$ and $R_1 \we R_2$. Thus the set $\EC (X)$ is a lattice, the operations $\we$ and $\vee$ being easily defined: the first one is simply the set-theoretical intersection  ($R_1 \we R_2 = R_1 \cap R_2$), the second one is obtained by taking, for any two elements in $\EC (X)$, the intersection of all $R\in \EC (X)$ containing both.  

Actually, the pair $(\EC (X),\preceq)$ is a {\sl complete} lattice, which means that for every nonempty subset $S$ the highest lower bound and the lowest upper bound of $S$ exist; they are denoted, respectively, by

\[ \bigwedge_{R\in S} R, \; \bigvee_{R\in S} R. \] 

A group $G$ acting on $X$ also acts naturally on $X\times X$ (componentwise), and this in turn induces an action on $\EC (X)$; more precisely, if $R\in \EC (X)$, the set $g\cdot R$ is also in $\EC (X)$ for all $g\in G$. Explicitly, if $p,q\in X$, and $g\in G$, then $g\cdot R$ is the element of $\EC (X)$ defined by 

\[ p (g\cdot R) q \Llr (g^{-1} \cdot p) R (g^{-1} \cdot q). \]

\begin{df} Let $G$ be a group acting to the left on the set $X$. An equivalence relation $\sim$ is {\bf G-invariant} if it is a fixed point of the induced action of $G$ on $\EC (X)$.\end{df}

\n
In words, $\sim$ is $G$-invariant if and only if for every $p,q\in X$ and for every $g\in G$, $p\sim q$ implies $g\cdot p\sim g\cdot q$. 

\begin{pro} The set $\EC (X)_G$ of all $G$-invariant equivalence relations on $X$ is a complete sublattice of $\EC (X)$.\end{pro}

\n
{\bf Proof}  It is enough to prove that for every $R_1 , R_2\in \EC (X)$ and $g\in G$, we have

\[ g\cdot (R_1 \vee R_2) = (g\cdot R_1) \vee (g\cdot R_2), \; g\cdot (R_1 \we R_2) = (g\cdot R_1) \we (g\cdot R_2). \]  

\n
The second identity follows immediately from the fact that the map defined by $g$ on $X\times X$ is a bijection, and therefore preserves set-thoretical intersection. Almost as immediate is the validity, with a similar justification, of the first identity. Completeness just follows from the fact that any intersection of $G$-invariant subsets in $X\times X$ is also $G$-invariant. \hfill$\Box$

Since the set of all $G$-invariant equivalences which are less fine than a given $R\in\EC (X)$ is nonempty (it comprises at least $T$), it has an infimum, which is actually a minimum; we shall call it the $G$-invariant equivalence {\sl generated}  by $R$, and will denote it by $\tl{R}$. It is easy to see that the following equality holds:

\be \tl{R} = \bigvee_{g\in G} g\cdot R .\ee{gen_er} 

Given a $G$-invariant equivalence relation $\sim$ we can define for every point $p$ the set

\[ \tlH (p) := \{g\in G \; : \; g\cdot p \sim p\} \]

\n
which is also easily shown to be a subgroup (the $\sim$-{\sl isotropy} subgroup of $p$), clearly satisfying:

\be H(p) \leq \tlH (p) \leq G. \ee{sgr}

\n
Also, for every $g\in G$ one has (compare with \rf{iso}):

\be \tlH (g\cdot p) = g\tlH (p)g^{-1}.\ee{sgr_v}

\subsection{Transitive and nontransitive actions}

Let us assume that the action of $G$ on $X$ is transitive. From \rf{sgr} and \rf{sgr_v} it follows that $\sim$ is trivial if either inclusion in \rf{sgr} is an equality for some $p\in X$. Clearly, in order for this to be true of {\sl all} $G$-invariant equivalence relations a {\sl sufficient} condition is that there are no subgroups $K$ satisfying the strict inclusions

\be H(p) < K < G. \ee{sgr_s}

\n
This in turn means that either $H(p) = G$ (which is equivalent to saying that $X$ is a singleton), or $H(p)$ is a {\sl maximal} subgroup (i.e. a proper subgroup not contained in any other proper subgroup of $G$).

This condition is also {\sl necessary}, since if a subgroup $K$ verifying \rf{sgr_s} exists, then for a fixed $p\in X$ one can define the equivalence relation:

\be g_1\cdot p \sim g_2\cdot p : \Llr g_1^{-1}g_2 \in K, \ee{rel_K}

\n
which is clearly $G$-invariant and satisfies $\tlH (p) = K$; its equivalence classes are the sets

\be [g\cdot p]:= (gK)\cdot p,\; \mbox{for all}\; g\in G \ee{eq_cl}

\n
This equivalence relation depends on the base point $p$ unless $K$ is a normal subgroup; however, because of \rf{sgr} and \rf{sgr_v}, all $G$-invariant equivalence relations on $X$ can be obtained this way for any fixed base point. So we can state:

\begin{pro} If $G$ is a group acting transitively on a set $X$, the nontrivial $G$-invariant equivalence relations on $X$ are parametrized by all subgroups $K$ satisfying \rf{sgr_s} for any fixed $p\in X$, the equivalence classes being given by \rf{eq_cl}.\hfill \hfill \hfill $\Box$     \end{pro}

On the other hand, if the action is not transitive, the classes of inequivalent points need not be of same size, thus an invariant equivalence relation is trivial if and only if  either inclusion in \rf{sgr} is an equality for all points. As to equivalence classes, each one is partitioned into subsets of the type $\tlH (p)\cdot p$. Thanks to their subgroup of translations the main groups we shall consider (the Newton, Galilei and Poincar\'e groups) {\sl all act transitively on $\R^4$}. However we shall have to deal also with the nontransitive actions of {\sl subgroups} of these groups.

\subsection{Semi-direct products}

In order to classify all invariant equivalence relations according to the Galilei and the Poincar\'e groups on $\R^4$ we shall need a simple proposition on semi-direct products of subgroups of a given group $\hG$.  We recall that $G = G_0 \sd  G_1$ means that $G_0$ and $G_1$ have trivial intersection, that their union generates the whole of $G$, and that $G_1$ is a normal subgroup in $G$. It is easy to prove that in this case every $g\in G$ can be expressed uniquely as $g = g_0 g_1$ with $g_i \in G_ i$ ($i = 0,1$).\footnote{Note, however, that to say that $G = G_0 \sd  G_1$ does not identify $G$ up to isomorphism, contrary to the case of {\sl direct} product.}

\begin{pro} All subgroups $K$ of $G = G_0 \sd  G_1$ containing $G_0$ are of the form $K = G_0 \sd L$ with $L \lhd K$.\end{pro}

\n
{\bf Proof} Let $L := K\cap G_1$. If $k = g_0 g_1$ with $g_i \in G_ i$ ($i = 0,1$), then $g_0^{-1} k = g_1$ belongs to both $K$ and $G_1$, so it lies in $L$. It follows that $K = <G_0 \cup L>$, and of course $G_0 \cap L =\{e\}$. Finally, if $g_0 \in G_0$, then

\[ g_0^{-1} L g_0 \sq g_0^{-1} G_1 g_0 = G_1 \]

\n
and since $G_0 \leq K$, it must be $g_0^{-1} L g_0 \sq K\cap G_1 =L$, which is enough to conclude that $L$ is a normal subgroup in $K$. \hfill $\Box$    

Consider now the case  of a subgroup of $Aff(\R^n)$, the group of the affinities of $\R^n$. We denote by $GL(n,\R)$ the subgroup of the homogeneous affinities, and by $T(\R^n)$ the subgroup of all translations of  $\R^n$. We suppose that $G = G_0 \sd L$, where $G_0$ is a subgroup of $GL (n\R)$ and $L$ is a subgroup of $T(\R^n)$. Every such $L$  can be identified with a subgroup $\LI$ of $\R^n$ in a natural way:

\[ \LI =\{b\in\R^n \; :\; T_b\in L\}. \]

For $L$ to be normal in $G = <G_0 \cup L>$ a necessary and sufficient condition is that, for every $b\in \LI$ and $B\in G_0$, we have
\( B^{-1} T_b  B \in L \),  that is, since $B^{-1} T_b B = T_{B^{-1} b}$, it must be that $B (\LI) \sq \LI$ for every $B\in G_0$. In other words, $\LI$ must be a $G_0$-{\sl invariant} subgroup of $\R^n$. Thus, by applying \rf{eq_cl} we get:

\begin{thm} Suppose that $G = G_0 \sd T(\R^n)$, with $G_0$ a subgroup of $GL(n,\R)$. There is a one-to-one correspondence between the $G$-invariant equivalence relations on $\R^n$ and the $G_0$-invariant subgroups $\LI$ of $\R^n$; for every such $\LI$ the equivalence class according to $\sim_{L}$ of any point $x\in\R^n$ is:

\[ [x]_L = x+\LI. \]\end{thm}

\n
{\bf Proof} From the proposition 2.4 and the remarks following it, we have that all subgroups $K$ of $G = G_0 \sd T(\R^n)$containing $G_0$, which of course is the isotropy subgroup of $o = (0,0,\dots,0)$, are of the form $K = G_0 \sd L$ where the corresponding $\LI$ is a $G_0$-invariant subgroup of $\R^n$. The claim follows from applying \rf{eq_cl} with base point $o$, and noting that 

\[ [x]_L = T_x [o]_L = x + [o]_L = x + L\cdot o = x +\LI , \] 

\n
as claimed. \hfill$\Box$

\section{Simultaneity and invariant equivalent relations}
 
We are now ready to obtain a complete classification of the Galilei-invariant and Poincar\'e-invariant equivalence relations. First, however, it will be useful to investigate the more basic case of Newton invariance.

\subsection{Newton-invariant simultaneity}

As is well known, classical mechanics obeys the principle of relativity with respect to the Galilei group, which we denote by $\GC$, that is the group of all transformations on $\R^4$ of the form

\be g_{(S,\Wy,b)}: \R^4 \ra \R^4, \; x \mapsto \left(\ba{cc} S & \Wy \\ \0^T & 1 \ea\right)x + b, \ee{gg}

\n
where $S\in SO(3)$, $\Wy\in\R^3$ and $b\in\R^4$. A subgroup of $\GC$ is the set of all spatial rotations, which we shall denote by $SO_4 (3)$ and whose elements can be identified with matrices of the type:

\[ \Si_S =\left(\ba{cc} S & \0 \\ \0^T & 1\ea\right), \;\mbox{with}\; S \in SO(3). \]

However, as is also well known, Newton's {\sl Principia} presents a theory based on the concepts of absolute space and time. We can say that this theory also obeys a principle of relativity, but here the structure group is a subgroup $\GC_N$ of $\GC$, which we will call the {\sl Newton group}, and whose elements are all the transformations of the type $g_{(S, \0,b)}$. The algebraic structure of $\GC_N$ as a subgroup of $Aff (\R^4)$ is given by:
  
\be \GC_N = SO_4 (3) \sd T(\R^4), \ee{new}
\n
that is, the Newton group is a semi-direct product of the space rotations and space-time translations. In order to find all the Newton-invariant equivalences, let us begin with a lemma:  

\begin{lem} Let $\by\in\R^3$ be any nonzero vector; then the additive subgroup of $\R^3$ generated by the subset $SO(3)\by =\{ S\by\; :\; S\in SO(3)\}$ is the whole of $\R^3$.\end{lem}

\n
{\bf Proof} Since $SO(3)\by$ comprises a (nonzero) vector for all directions, it is enough to show that the generated subgroup contains the whole half-line of the vectors positively proportional to $\by$. Let $\by_0$ be a vector orthogonal to $\by$ with the same module, and let $S_\te$ be the one-parameter subgroup of all rotations from $\by$ to $\by_0$ which fix the vector product $\by\we\by_0$. A simple computation (or an elementary geometric argument) shows that 
\((S_{\te} + S_{-\te})\by \) is proportional to $\by$ and has module equal to $2|\by|\cos\te$. Thus as $\te$ ranges over the interval $[0,\pi/2[$, we obtain all vectors with modules decreasing from $2|\by|$ to 0 having the same direction as $\by$. By summing all pairs of these vectors we obviously get the whole positive half-line, as claimed. \hfill \hfill $\Box$  

In the following statement and proof we use the symbols introduced in \S 2.4. 

\begin{pro} The $SO_4 (3)$-invariant subgroups $\LI$ of $\R^4$  are all and only those of either the form $\LI = \{\0\}\times H$ or the form $\LI = \R^3 \times H$, where $H$ ranges over all subgroups of $\R$.\end{pro}

\n
{\bf Proof} That all subgroups of the type described are $SO_4(3)$-invariant is clear. Conversely, suppose $\LI\neq\{\0\}$ and define \( H = \{h\in\R\; :\; (\0,h)\in \LI\} \). Clearly $H$ is a subgroup of $\R$.  If all $a =(\ay, a^4)$ in $\LI$ have $\ay = \0$, then obviously $\LI = \{\0\}\times H$. If there is at least an $a$ with $\ay\neq \0$, then let $S$ be an arbitrary rotation which does not fix $\ay$; thus the following vector:

\[ \Si_S a -a = (S\ay -\ay , 0) \]

\n
belongs to $\LI$, and therefore, using the lemma 3.1 for $\by: = S\ay-\ay$, we get that $\LI$ contains $\R^3 \times \{0\}$, and therefore also $\R^3 \times H$; for the reverse inclusion note that if $a\in \LI$, then $(\ay, 0)$ and $a-(\ay,0)$ are also in $\LI$, so $a^4\in H$: it follows that $a\in \R^3\times H$. \hfill $\Box$        

From theorem 2.5 and proposition 3.2 we deduce: 

\begin{thm} The $\GC_N$-invariant equivalence relations on $\R^4$ are of two types, each one depending on the arbitrary choice of an additive subgroup $H$ of $\R$:

(i)  $ [x] = x +(\{\0\}\times H)$,

(ii) $ [x] = x + (\R^3 \times H) $. \hfill $\Box$  \end{thm}

Now I recall the following elementary result, for which a proof is provided for the reader's commodity in the Appendix; remember that a topological space is called {\sl totally disconnected} if all its connected components are singletons.

\begin{lem} All proper subgroups $H$ of $\R$ are totally disconnected; they are either of the form $\Z a$, with $a\in \R^+$, or everywhere dense.\end{lem}

\n
Examples of subgroups of the second type (that is, everywhere dense subgroups)  are the rational numbers $\Q$ or the sets $\Z + \Z a$, for irrational $a$.

``Time'' as defined by a $\GC_N$-invariant equivalence relation $\sim$ is represented by the quotient space $\R^4/\sim$ -- the {\sl space of instants}. From theorem 3.3 there are two types of such quotients:

\[ \R^4 / (\{\0\}\times H) \cong \R^3 \times (\R/H), \;  \R^4 / (\R^3\times H) \cong \R/H .\]

\n
Let us remark that the second type, although closer to the intuitive idea of time, gives rise nonetheless to several non-homeomorphic topological spaces: apart from 1) a singleton (if $H =\R$), and 2) the real line (if $H=\{0\}$), we also have 3) the circle (if $H = \Z a$) and 4) an {\sl infinite indiscrete space} (if $H$ is an everywhere dense subgroup). 

As is clear from these results, even in Newton space-time the invariant equivalence approach is far from giving us an uniqueness theorem. To obtain a reasonable $\GC_N$-invariant concept of simultaneity something more is needed. There are essentially two ways to proceed: either we require, additionally, that the equivalence classes enjoy some further property; or, alternatively, we enforce invariance with respect to a suitably bigger group. In the latter case, one may introduce the {\sl conformal Newton group}        

\be C\GC_N := \R^+ SO_4 (3) \sd T(\R^4), \ee{new_c}

\n
which corresponds to the assumption that there are no absolute time and space scales, {\sl but that their ratio is fixed}. The following proposition lists two options to achieve uniqueness.  

\begin{pro} Standard Newtonian simultaneity is the only nontrivial $\GC_N$-invariant equivalence relation on $\R^4$ such that either (i) its classes are connected subspaces; or (ii) it is $C\GC_N$-invariant and its equivalence classes are not worldlines. \end{pro}

\n
{\bf Proof} (i) This follows from the fact that, by the lemma 3.4, the only connected subgroups of $\R$ are the trivial ones. (ii) A $C\GC_N$-invariant equivalence is obviously also $\GC_N$-invariant, thus we must check which nonzero subgroups $\LI$ of $\R^4$ in proposition 3.2 turn out to be also $C\GC_N$-invariant. Now for $H\neq \{0\}$

\[ \R^+ (\{\0\}\times H) = \{\0\}\times \R,\] 

\n
and this means that the equivalence classes are worldlines; in the other case \(\R^+ (\R^3\times H)\) is equal to $\R^4$ (total equivalence) unless $H=\{0\}$, which gives the standard simultaneity, as claimed. \hfill $\Box$

\subsection{Galilei-invariant simultaneity}

Since the Newton group $\GC_N$ is a subgroup of the Galilei group $\GC$, the $\GC$-invariant equivalence relations form a subset of the $\GC_N$-invariant equivalence relations. Thus by using theorem 3.3 to select those $\GC_N$-invariant relations which happen to be also $\GC$-invariant, we can state the full theorem on $\GC$-invariant equivalence relations on $\R^4$ (which corrects ``Theorem 2'' in \cite{g}):\footnote{The argument in \cite{g}, p. 661, fails to consider the `everywhere dense' case of our lemma 3.4.}

\begin{thm} All nontrivial $\GC$-invariant equivalence relations on $\R^4$ are parametrized by proper subgroups $H$ of $\R$; the equivalence classes of $\sim_H$ are of the form

\[ [x]_H = x + (\R^3\times H). \]\end{thm}

\n
{\bf Proof} In fact uniform motions in $\GC$ discard case (i) in Theorem 3.3 . \hfill $\Box$ 

Since the only proper subgroup $H$ of $\R$ which is connected is $\{0\}$ (cf. lemma 3.4), we have, by introducing the (naturally defined) conformal Galilei group $C\GC $: 

\begin{pro} Standard Newtonian simultaneity is the only nontrivial $\GC$-invariant equivalence relation on $\R^4$ such that either (i) its classes are connected subspaces; or (ii) it is $C\GC$-invariant. \hfill $\Box$  \end{pro}

\subsection{Poincar\'e-invariant simultaneity}

We denote by $\LC$ the Lorentz group on $\R^4$, by $\LCU$ its orthochronous (i.e. time-orientation preserving) subgroup, and by $\PC$ (resp. $\PCU$) the Poincar\'e (the orthochronous) group, that is:

\[ \PC = \LC \sd T(\R^4), \; \PCU = \LCU \sd T(\R^4). \]  

\n
The {\sl proper}  groups are the intersections of these groups with the orientation-preserving affinities of $\R^4$: $\LC^+, \LCU^+, \PC^+, \PCU^+$. Nothing essential is lost from using the proper subgroups rather than the corresponding larger groups, so to comply with tradition and for consistency with the treatment of the classical cases we shall use the former in the following.  

By $g_\la$ for any $\la>0$ we denote the Lorentzian form on $\R^4$ represented in the canonical basis by the diagonal matrix with entries $(1,1,1,-\la^2)$. The form to which $\LCU^+$ refers is $g \equiv g_c$, where $c$ is the speed of light in empty space.

From an algebraic point of view the treatment of Poincar\'e-invariant simultaneity parallels what we have just done, since

\be \PCU^+ = \LCU^+ \sd T(\R^4). \ee{poi}

\n
Therefore, from theorem 2.5 we know that all Poincar\'e-invariant equivalence relations are parametrized by $\LCU^+$-invariant (additive) subgroups $\LI$ of $\R^4$. The search for interesting invariant equivalence relations is stopped by the following:

\begin{pro} There are no nonzero $\LCU^+$-invariant proper subgroups $\LI$ of $\R^4$.\end{pro}

\n
{\bf Proof} Since $SO_4 (3)$ is contained in $\LCU^+$, we can use proposition 3.2 and look for those nonzero subgroups of the type $\LI = \R^3 \times H$ which are also invariant for $\LCU^+$. To get our result, it suffices to remark that if $\La\in\LCU^+\setminus SO_4 (3)$ (say, let $\La$ be a Lorentz boost, i. e. a special Lorentz transformation with nonzero velocity), then $\La (\R^3 \times H)$ contains an hyperplane $A$ which is not parallel to $\R^3 \times \{0\}$. If we call $\pi$ the natural projection of $\R^4$ on the fourth factor (i.e. $\pi (x^1, x^2, x^3, x^4) = x^4$), then for any such hyperplane one has $\pi (A) =\R$ and since $\pi (\R^3 \times H) = H$, it must be $H=\R$.\hfill \hfill $\Box$  

\begin{cor} There are no nontrivial $\PCU^+$-invariant equivalence relations on $\R^4$.\hfill \hfill $\Box$  \end{cor}

Of course {\sl a fortiori}\footnote{This is theorem 2 in \cite{ho05}, p. 496, which, however, contains an evident material error in its thesis.} there are no nontrivial $\PC$-invariant or $\R^+\PC$-invariant equivalence relations on $\R^4$. 

Here is a simpler result, which follows from the previous ones, but that it is instructive to prove independently. Let us call $R_u$, for every future timelike $u$, the standard simultaneity relation determined by $u$, that is:

\be x R_u y \Llr x-y\perp u . \ee{ssr}

We can ask what is the $\PCU^+$-invariant relation generated by $R_u$. Since for every $g\in \PCU^+$ with linear part equal to $\La\in \LCU^+$ we have \( g\cdot R_u = R_{\La u}\), from \rf{gen_er} we infer:

\[ \tl{R_u} = \bigvee_{v\in\TC^+} R_v = T .\]
 
\n
where $\TC^+$ is the set of all future timelike free vectors. The last equality is the statement of the geometrically intuitive fact that the only equivalence relation such that every event is equivalent to {\sl at least} all events with space-like separation from it is the total one. This result can be much refined. In fact:

\begin{pro} If $u_1$ and $u_2$ are non-proportional future timelike vectors we have 

\[ R_{u_1}\vee R_{u_2} = T. \]\end{pro}

\n
{\bf Proof} Let us denote $R_{u_1}\vee R_{u_2}$ by $\sim$. It is enough to prove that for a given $x\in \R^4$ and every $s\neq 0$, $x\sim x+s u_1$, which follows from

\[ (x+<u_1>^\perp ) \cap (x+ s u_1 + <u_2>^\perp) \neq \emptyset ,\]

\n
where $<a>$ is the vector subspace generated by $a\in \R^4$. Now this condition means that there is $y$ such that 

\[ g (y-x, u_1) =0, \; g (y-x, u_2) = s g(u_1, u_2) .\]

\n
Since $u_1$ and $u_2$ are not proportional it is easy to see that this system of two linear equations in $y$ always has a solution. \hfill \hfill $\Box$  

This proposition can be expressed in words by saying that the only equivalence relation which is compatible with the synchronies of just (any) two inertial observers which are not mutually at rest is the total one. On the other hand the only equivalence relation resulting from the requirement that couples of events are related if they are synchronous according to just (any) four linearly independent inertial observers $v_1, v_2, v_3, v_4$ is the identical one:

\[ R_{v_1} \we R_{v_3} \we R_{v_3} \we R_{v_4} = \bigwedge_{v\in\TC^+} R_v = I .\]

\n
This shows that there is no easy escape from corollary 3.9: the Poincar\'e-invariant equivalence relation approach leaves us with the choice between an Absolute Present shrinking to a single event and an Absolute Present expanding to include the whole space-time.\footnote{The results of this section form the technical core of the debate on the so-called block-universe concept in special relativity. I will deal with it in a future article.}

\subsection{Causality and simultaneity}

We have seen that the invariance under the structure group is not enough to select a unique nontrivial invariant equivalence relation, either in classical (Galilei and Newton) or in Minkowski spacetimes. Topological constraints on the equivalence classes, as we have seen, provide standard simultaneity in the classical cases. A different approach is to look at whether a natural causality condition is satisfied.  

The so-called ``causal theory of simultaneity'', as endorsed by H. Reichenbach and A. Gr\"unbaum (cf. recently \cite{gru}), postulates that events which can be possibly in a causal relation, given the causal structure of the relevant space-time, cannot be simultaneous, and that {\sl this is the only nonconventional constraint on simultaneity}. It is not my purpose in this paper to discuss this theory; I have argued at length in \cite{mm}  for a different way of grounding physically the concept of simultaneity. In the following the {\sl causality condition}  on simultaneity is that causally connectible events cannot be simultaneous. Of course which pairs of events are causally connectible depends on the (causal structure of the) space-time.

In the Newton and Galilei space-times (i.e. the space-times modelled on $\R^4$ as a $\GC_N$-space and, respectively, as  a $\GC$-space), any two events $p$ and $q$ are causally connectible if, given an admissible time function $t$, one has $t(p) \neq t(q)$. It follows that both cases (i) and (ii) in Theorem 3.3 violate the causality condition in {\sl Newton space-time} unless $H$ is the zero subgroup; in this case the only nontrivial equivalence turns out to be (ii), and it coincides with Newtonian simultaneity. We can state it formally as:

\begin{pro}  The only $\GC_N$-invariant (resp. $\GC$-invariant) equivalence relation on $\R^4$ satisfying the causality condition is absolute simultaneity. \hfill $\Box$ \end{pro}

In Minkowski space-time, if we denote by $\TC$ (resp. $C$) the set of all timelike free vectors (resp. the lightcone\footnote{I.e., the set of all lightlike vectors, which does not comprise the zero vector (cf. \cite{sw77}, p. 22).} together with its vertex), then $p, q$ are {\sl causally connectible} if and only if $p-q$ lies in $\TC\cup \ol{C}$. If $p$ and $q$ are causally connectible and $p$ chronologically precedes $q$, we write $p \leq q$.   

The next theorem relates the two-way light isotropy, causality and simultaneity. I will first need two definitions and a proposition.

\begin{df} An {\bf inertial coordinate system} in Minkowski space-time is an affine coordinate system such that the associated basis of free vectors is of the form $(e_1, e_2, e_3, u)$, where the generated subspace $[e_1, e_2, e_3]$ is spacelike, and $u$ is timelike.\end{df} 

\begin{pro} In Minkowski space-time, for every inertial coordinate system $\phi'$ such that the isotropy of the two-way velocity of light holds, there exists a unique Minkowski coordinate system $\phi$ such that the transition function from $\phi$ to $\phi'$ is of the form:

\be \ry' = \la \ry,\; t' = \la (t + \ky \cdot\ry), \; \mbox{with}\; \la >0,\; |\ky |<1/c . \ee{rei}

\n
Vice versa, every $\phi'$ related to a Minkowskian $\phi$ by a transition function of the form \rf{rei} satisfies the two-way isotropy of the velocity of light.  \end{pro}

\n
{\bf Proof} This is an easy consequence of the theorem at p. 795 and the proposition 1 at p. 797 of \cite{mm}. \hfill $\Box$   

\begin{df} Let $\phi$ be any inertial coordinate system on $\R^4$, and $p\neq q$ events in $\R^4$; we say that $p$ is {\bf causally M-connectible} to $q$ {\bf according to $\phi$} if $\phi (q) - \phi(p)$ is a future timelike or null vector. Every statement to the effect that an event is causally M-connectible to another event will be called a {\bf causality M-statement}.
\end{df}

\begin{thm} Let $\phi$ be a Minkowski coordinate system, and suppose that $\phi'$ is an inertial coordinate system which is at rest with respect to $\phi$ and for which the isotropy of the two-way velocity of light holds. If every causality M-statement which holds true according to $\phi$ also holds true according to $\phi'$, then $\phi'$ is a Minkowski coordinate system up to a scale factor, and the $\phi$-synchrony and the $\phi'$-synchrony coincide.\end{thm} 

\n
{\bf Proof} From proposition 3.13 above it follows that 

\[ \ry' = \la A\ry,\; t' = \la (t + \ky\cdot A\ry), \; \mbox{with}\; \la >0,\; |\ky|<1/c, \]

\n 
$A$ being an orthogonal 3x3 matrix. By assumption, for every 3-vector $\vy$ such that $|\vy|\leq c$, the following inequality holds:

\be 0\geq q(g)(x'(t\vy, t)-x'(\0,0)) = \la^2 t^2 (|\vy|^2 -c^2 (1+\ky\cdot A\vy)^2) \ee{co_ca} 

\n
where $x'=(\ry', t')$. It follows that

\[ |\vy|\leq c (1+\ky\cdot A\vy). \]

\n
Now, if $\ky\neq \0$, a choice of $\vy$ with direction opposite to that of $A^T \ky$ and with module equal to $c$ would clearly violate this inequality. Thus $\ky = \0$, as required. \hfill \hfill $\Box$  

From this proposition it might be argued that the mere requirement of causality, added to the two-way light isotropy condition, is enough to single out standard synchrony, thus defeating one of the conventionalist claims. In \cite{m77} D. Malament attributed to A. Gr\"{u}nbaum the claim that standard synchrony ``is not uniquely definable in terms of the relation of causal connectibility'' and countered it by stating instead that ``the relative simultaneity relation of special relativity {\sl is} uniquely definable from the causal connectibility relation''. There are other results, by A. D. Alexandrov \cite{a75} (partly rediscovered by E. C. Zeeman), which establish another strong link between causality and the conformal Poincar\'e group. One of them can be stated as follows, where we denote by $p\leq q$ the condition that $q-p$ is either zero or a future-pointing causal vector:

 \begin{thm} [Alexandrov] Let $f: \R^4 \ra \R^4$ be a bijection, with $\R^4$ endowed with the standard orthochronous Lorentz structure, such that for every $p,q\in \R^4$

\[ p \leq q  \Longleftrightarrow f(p) \leq f(q), \]

\n
then $f$ is a conformal orthochronous Poincar\'e transformation.\end{thm} 

\n
We shall deal in the final section with the question whether these results provide a `causal' justification of standard simultaneity.

\subsection{The isotropy subgroup of a notion of rest}

Given proposition 3.8, if we look for nontrivial invariant equivalence relations, clearly we fare no better by taking a {\sl larger} group than $\PCU^+$. On the other hand, if we take a {\sl smaller} group nontrivial equivalence relations may be found.

D. Giulini (\cite{g}) introduces the subgroup $(\PCU^+)_X$ of all Poincar\'e transformations fixing a given ``inertial frame'', which is a complete family of parallel timelike worldlines; in the notation of \cite{mm}, $X =\G (u)$, where $u$ is a future timelike vector giving the direction of the pencil.

Clearly if we define \( H(u) := \{ \La \in \LCU^+\; :\; \La u = u\} \) we have

\[ (\PCU^+)_X = H(u) \sd T(\R^4). \]

Application of theorem 2.5 leads to the search for all $H(u)$-invariant additive subgroups $\LI$ of $\R^4$. Since for every nonzero scalar $\la$ and for every $\La\in\LCU^+$ one has

\be H(\la u) = H(u), \; H(\La u) =\La H(u)\La^{-1}, \ee{isot}

\n
we can start by taking $u=e_4$, thus reducing the problem to Newton invariance, since \( H(e_4) = SO_4 (3)\). By using proposition 3.2 we have:

\[ \LI = \{\0\}\times H \; \mbox{or}\; \LI = \R^3\times H, \]

\n
where $H$ is any additive subgroup of $\R$. The identities \rf{isot} allow us to infer, for any timelike $u$, the following result, which reformulates and corrects ``Theorem 5'' in \cite{g}; part B) considers the conformal case. We denote by $CH(u)$ the isotropy subgroup in $\R^+\LCU^+$ of the one-dimensional vector space $<u>$  generated by $u$. It is easy to see that $CH(e_4) = \R^+ SO_4 (3)$. We define 

\[ C(\PCU^+)_X := CH (u) \sd T(\R^4). \]   

\begin{thm} A) For every timelike vector $u$, the $(\PCU^+)_X$-invariant equivalence relations are of two types, each one depending on the arbitrary choice of an additive subgroup $H$ of $\R$:

(i)  $ [x] = x +Hu $,

(ii) $ [x] = x +(<u>^\perp + Hu) $. 

B) Standard synchrony $R_u$ is the only  $C(\PCU^+)_X$-invariant equivalence relation whose equivalence classes are not worldlines.
 \end{thm}

\n
{\bf Proof} (A) From \rf{isot}, it follows that if $\La e_4 = u$, then $\LI$ is a $H(u)$-invariant subgroup of $\R^4$ if and only if 
\(\La^{-1} \LI \) is $SO_4 (3)$-invariant. By using theorem 3.3 both (i) and (ii) follow.

(B)  Just apply point (ii) of proposition 3.5. \hfill \hfill $\Box$  

An immediate consequence of this theorem is:

\begin{cor} If the classes of a $(\PCU^+)_X$-invariant equivalence relation are 3-dimensional manifolds, or connected topological spaces with more than one point, the only possible synchrony is the standard one. \hfill $\Box$ \end{cor}

\subsection{The isotropy subgroup of a timelike straightline} 

Actually Malament (\cite{m77}, p. 297) stated in 1977 a result which looks more impressive than theorem 3.17(B) or corollary 3.18, and which has provoked much discussion, also arousing doubts about its correctness. I will now state it and prove it in a simple and geometrically transparent way which exploits the techniques which have been presented in this paper and includes the extension made in \cite{ss99}. The details I give will make it easier, I hope, to understand its exact (and unfortunately rather limited) scope. 

First a few definitions are in order. Let us call $\Te$ the time inversion map on $\R^4$:

\[ \Te: \R^4 \ra \R^4, \; x=(\x, t)\mapsto (\x, -t), \]

\n
and denote by $D$ the group generated by $\Te$ (clearly $D = \{\Te,I\}$). 

Let us call $\ell$ any timelike inertial worldline in $\R^4$, with normalized (i.e. with norm equal to $c$) future vector $u$. The standard simultaneity associated to $\ell$ is just $R_u$, as defined in \rf{ssr}. The group of the causal automorphisms {\sl plus} time inversion is the following:

\[ C\PC = \R^+ \LC \sd T(\R^4). \]

\n
We denote the isotropy subgroup of $\ell$ in $\PC$ (resp. in $\GC_N$) by $G_M$ (resp. $G_N$), and by $\hG_M$ (resp. $\hG_N$) the isotropy subgroup of $\ell$ in $C\PC$ (resp. $C\GC_N$). Moreover we define for all $\hc>0$

\[ C^+_{\hc} = \{ x\in \R^4 \; : \;  g_{\hc} (x,x) = 0, x^4 >0\}, \; C^-_{\hc} = \{ x\in \R^4 \; : \;  g_{\hc} (x,x) = 0, x^4 <0\}.\]

\begin{thm} [\cite{m77}, \cite{ss99}]  Let $\sim$ be a nontotal equivalence relation on $\R^4$ such that the equivalence class of some point in $\ell$ is not contained in $\ell$. 

1) If $\sim$ is $\hG_N$-invariant, then $\sim$ is either $R_u$ or there is a $\hc>0$ such that for every $p\in\ell$, the equivalence class of $p$ is 

\[ [p] = p + X\]

\n
where $X$ is $C^+_{\hc} \cup \{p\}$ (resp. $C^-_{\hc} \cup \{p\}$), that is, the equivalence classes for points of $\ell$ are translated half-cones along $\ell$.  

2) If $\sim$ is $\hG_M$-invariant, then $\sim = R_u$.\end{thm}

\n
{\bf Proof} 1) With no loss of generality, we can take $\ell$ to be the set $o+\R  e_4$, that is the ``vertical'' line passing through the origin $o$ of $\R^4$ in the usual space-time diagrams. With this choice we have

\[ \hG_N = \R^+ SO_4 (3)\sd \R e_4. \]

\n
Let us call $a = (\ay, a^4)$ a point in $[o]$ which is not in $\ell$; this implies that $\ay\neq \0$. For every $S\in SO(3)$ and $\la\in \R^+$ we have \( o\sim \la \Si_S a = (\la S\ay, \la a^4) \), whence it follows for every $k\in\R$ that

\[ [(\0, k)] = (\0, k) + [o] \sqq [(\la S \ay, k +\la a^4)].\]

\n
It is easy to see that for any $x = (\x, x^4)$ outside of $\ell$  (that is, $\x \neq \0$), the system of equations in the unknowns $\la >0$, $S\in SO(3)$ and $k\in\R$,

\[ \la S \ay = \x, \; k +\la a^4 = x^4 \]   
                       
\n
can always be solved; therefore \( \bigcup_{k\in\R} [(\0, k)] =\R^4\), that is, the equivalence classes of all points in $\ell$ are {\sl all} the classes in the quotient set $\R^4 /\sim$. 

Moreover, no two points in $\ell$ can be equivalent. It is enough to show it for $o$ and any $(\0, k)$, with $k\neq 0$. If such an equivalence held, then for every $\la >0$ \( [o] = \la [o] = [(\0,\la k)] \), which means that $o$ is equivalent to the whole half-line of $\ell$ to which $(\0, k)$ belongs. On the other hand, 

\[ (\0, -k) = T_{-ke_4}  (o) \sim T_{-ke_4}  (\0, k)  = o \]

\n
thus {\sl all}  points in $\ell$  would be equivalent and $\sim$  would be the total equivalence, which it cannot be. 

Now, suppose that $a^4 >0$ (the other inequality can be taken care of by a completely analogous argument). The set of all $(\la S\ay, \la a^4)$, which is contained, as we have seen, in $[o]$, is the upper half of the cone quadric

\[ |\x|^2 - \hat{c}^2 (x^4)^2 = 0, \; x^4 >0 \]

\n
where $\hat{c}: = |\ay /a^4|$. Therefore, since $((\0, k) + (C^+_{\hc}\cup \{o\}))_{k\in\R}$ is for any $\hc>0$ a partition of $\R^4$, the strict inclusion

\[ [(\0, k_1)] \supset   (\0, k_1) + (C^+_{\hc}\cup \{o\}) \]

\n
for some $k_1$  would imply that $(\0, k_1) \sim (\0, k_2)$  for some $k_2 \neq k_1$, which is absurd. 

2) With the same choice for $\ell$, 

\[ \hG_M = \R^+ (SO_4 (3)\sd D)\sd \R e_4. \]

\n
If $\sim \neq R_u$, then since \(C^-_{\hat{c}} = \Te (C^+_{\hat{c}})\) we obtain that

\[ [(\0, k)] = (\0, k) +(\{o\}\cup C^+_{\hat{c}}\cup C^-_{\hat{c}}). \]

\n
From this inclusion, however, it follows that all points in $\ell$ are equivalent and therefore, absurdly, that $\sim$ is the total equivalence. This consequence is clear geometrically; formally, if $k_1 < k_2$, then every point $x = (\x, (k_1 + k_2)/2)$ such that 

\[ |\x|^2 = \hc^2 (\frac{k_2 - k_1}{2})^2 \]

\n
belongs to the intersection of $(\0, k_1) + C^+_{\hat{c}}$ and $(\0, k_2) + C^-_{\hat{c}}$, which implies that $(\0, k_1) \sim (\0, k_2)$, contrary to what we have seen above.  \hfill \hfill $\Box$  

\section{Discussion}

According to the supporters of the causal theory of simultaneity (CTS) any condition on simultaneity in addition to the causality condition is for this very reason conventional, and this {\sl includes even such a formal, basic condition as that simultaneity must be an equivalence relation}. I consider this formal condition on the same foot as the condition of symmetry on the ordinary spatial distance (see \cite{mm}, pp. 784-6), thus in my view a sound methodological argument in favour of a synchronization procedure is that it issues into an equivalence relation, and this in turn implies that the CTS must be rejected.  On the other hand, adherence to the CTS implies that only where absolute simultaneity holds (i.e. in classical space-times) simultaneity happens to be an equivalence relation. So the whole issue of simultaneity as an {\sl invariant} equivalence relation is strictly speaking irrelevant (or question-begging) with respect to the debate on the epistemological status of simultaneity, as long as conventionalists rest their case on the CTS.\footnote{What amounts essentially to the same criticism has also been made in Gr\"unbaum's reply (\cite{gru}, p. 1288).} In the rest of this section I shall review the attempts at getting around this stricture which have been described in mathematical terms in the previous sections.  

We have seen in section 3.4 that the uniqueness of simultaneity holds if we enforce ``causal M-connectibility'' instead of ``causal connectibility''. On the other hand,  causal connectibility, as the (absolute) possibility of a causal interaction between events, is equivalent to the (relative) ``causal M-connectibility'' {\sl only
 in Minkowski coordinate systems}. In any other inertial system, {\sl there are causally connectible pairs of events which are not causally M-connectible according to that system}. In short, to define a causality statement in terms of the standard light-cone in $\R^4$, that is, as a causality M-statement, preempts the conventionality issue. In non-Minkowskian inertial systems causality statements just cannot be construed as statements about where the differences between 4-tuples lie with respect to the Minkowski light-cone of coordinate space. To be more explicit, by reference to proposition 3.13, the light cone of $\phi$ at the origin

\[ |\ry|^2 -c^2 t^2 =0\]

\n
corresponds in $\phi'$  to

\[ |\ry'|^2 -c^2 (t'-\ky\cdot \ry')^2 = 0,\] 

\n
which of course is still a cone in $\R^4$, but not the standard light-cone. Physically speaking, this simple geometric fact means that in a non-Minkowskian inertial system there are signals (i.e. physical  processes which causally connect events) which are one-way faster than $c$. But this should cause neither surprise nor discomfort to the conventionalists.

What about the reasonable refinement obtained by considering not the whole Poincar\'e group, but the isotropy subgroup of a notion of rest? Does this move adjudicate the debate on conventionality of simultaneity to one side or the other? 

The answer is a sobering no. Let us simplify this approach as asking which partitions of $\R^4$ into hyperplanes are preserved by those Poincar\'e transformations whose linear part has $u$ as an eigenvector.\footnote{Compare \cite{mm11}, pp. 1382-3.} Clearly, any Lorentz map which preserves (the direction of) $u$ must also preserve its Lorentzian {\sl orthogonal} complement, and the set of all Lorentz maps which preserve $u$ can preserve no other vector hyperplane. This fact, which we have proven above in a more general setting (\S 3.5), and is also intuitive in this simplified formulation, indicates where the trap is. Those who reject the uniqueness of synchrony relative to $\G (u)$, {\sl also reject the link between the physical evidence underlying special relativity and the Poincar\'e group}. In other words, according to their view, the Poincar\'e group, insofar as it embodies the assumption of one-way isotropy for the velocity of light, {\sl goes beyond physical evidence}, and therefore is not sacrosanct. It follows that any defence of standard synchrony using invariance under some subgroup of the Poincar\'e group simply begs the question. 

The same irrelevance applies, with a vengeance, to Malament's result.  I say `with a vengeance' since the proper environment in which it makes sense to compare different synchronies is given not by a single inertial wordline, but by a pencil $\G (u)$ -- our notion of rest. However, the introduction of space-time dilatations and, worse, of time inversion, seems totally unwarranted from a physical point of view. Hogart (\cite{ho05}) has qualms about dilatations and proves a proposition relying on $G_M$ rather than on $\hG_M$, and which in our terms can be reformulated and proved as follows:

\begin{pro} [\cite{ho05}] Let $\sim$ be an equivalence relation on $\R^4$ such that all equivalence classes have a unique representative in $\ell$.   
If $\sim$ is $G_M$-invariant, then $\sim = R_u$.\end{pro}

\n
{\bf Proof} The hypothesis implies that $([p])_{p\in\ell}$ is a partition of $\R^4$, and the $G_M$-invariance implies that for every $p\in \ell$  
we have \( [p] = p+[o] \). Thus it is enough to prove that $[o] \sq o + [u]^\perp$. Let us consider, without loss of generality, $u=e_4$, and suppose by contradiction that there is a $p = (\ay, s) \sim o$, with $s\neq 0$. Then

\[ (\ay, 0) = T_{-se_4} (p) \sim T_{-se_4} (o) = (\0, -s). \]

\n
Since $[o]$ is $D$-invariant, it is also \( (\ay, -s) \sim o\), and by the same argument we have \(( \ay, 0) \sim (\0, s)\). Therefore $(\0, s) \sim (\0, -s)$, which would contradict the assumption of inequivalence of all points in $\ell$, unless $s=0$, as required. \hfill \hfill $\Box$   

\n
Even this proposition assumes time inversion, thus maintaining one of the more questionable aspects of Malament's result. As explained in theorem 3.19, 1), and as previously pointed out by Sarkar and Stachel (\cite{ss99}), if time inversion is suppressed within the single inertial line approach, a lot more invariant equivalence relations pop up.  In a sense, these are subtleties, as a double original sin is common to all of these failed attempts at refuting conventionalism: 1) assuming that simultaneity must be an equivalence relation, 2) relying on the Poincar\'e group and therefore, ultimately, on the one-way isotropy of light.

\section{Appendix}

We prove lemma 3.4, that is: {\sl All proper subgroups $H$ of $\R$ are totally disconnected; they are either of the form $\Z a$, with $a\in \R^+$, or everywhere dense}.

\n
{\bf Proof} It is enough to argue by restricting ourselves to the positive elements of $\R$. If $H$ is nonzero, then $H\cap\R^+$ is nonempty; let $a$ be its infimum. First suppose that $a$ is nonzero. If  $a$ does not belong to $H$, then for positive $\ep<a$ there exist $h_1, h_2\in H$ such that  \( a<h_1 <h_2 <a+\ep \), whence $h_2 -h_1 <\ep<a$ is a positive element of $H$ less than $a$, which is absurd. It follows that $a$ is the positive minimum of  $H$, so for every other $b\in H\cap \R^+$ it must be

\[ b = ma +r, \; \mbox{where}\; m\in \Z^+, 0\leq r <a. \]

\n
It follows that $b-ma$ is an element of $H\cap \R^+_0$ which is smaller than $a$, so it must be zero, and therefore $b\in\Z a$, that is $H = \Z a$.

Suppose now $a=0$; then there is a strictly decreasing sequence $(h_n)$ of elements in $H$ which converges to zero. Let $x$ be any positive real number. For every positive $\ep$, no matter how small, there is $n_0$ such that $h_{n_0} < \ep$. Thus by dividing $x$ by $h_{n_0}$ we get, for a suitable $n\in\Z^+$ and $r \geq 0$ strictly smaller than $h_{n_0}$ : \( x = nh_{n_0} + r\), and therefore $x - nh_{n_0} =r<\ep$, where of course $nh_{n_0}$ belongs to $H$. Thus in every neighborhood of every (positive) real number there lies some element of $H$, which means that $H$ is everywhere dense in $\R$. 

As to the topological structure, it is clear that every subgroup of the form $\Z a$ is totally disconnected. Suppose now that $H$ is everywhere dense; if the connected component $K$ of $0$ in $H$ is not a singleton, then it must contain an interval $[0,\ep[$ for some $\ep>0$, but then it must also contain $]-\ep, \ep[$ and therefore

\[ K\sqq \bigcup_{n\in \N} n]-\ep,\ep[ = \bigcup_{n\in \N} ]-n\ep, n \ep[ = \R, \] 

\n
which contradicts the hypothesis that $H$ is proper. \hfill \hfill $\Box$

\end{document}